\documentclass[aps,pre,showpacs]{revtex4}
\usepackage{graphicx}
\usepackage{bm}
\usepackage{amsmath}
\usepackage[centerlast]{subfigure}

\newif\iffigs
\figstrue   % uncomment if figures are present
\iffigs
\fi
\def\drawing #1 #2 #3 {
\begin{center}
\setlength{\unitlength}{1mm}
\begin{picture}(#1,#2)(0,0)
\put(0,0){\framebox(#1,#2){#3}}
\end{picture}
\end{center}}

\begin{document}
\title{Low frequency fluctuations in a Vertical Cavity Lasers:\\
experiments versus Lang-Kobayashi dynamics}
\author{Alessandro Torcini$^{(a,d)}$, Stephane Barland$^{(b)}$, Giovanni Giacomelli$^{(a)}$, and Francesco Marin$^{(c,d)}$}
\affiliation{
$(a)$ Istituto dei Sistemi Complessi - CNR, via Madonna del Piano 10, 50019 Sesto Fiorentino, Italy \\
$(b)$ Institut Non Lin{\'e}aire de Nice Sophia Antipolis and UMR 6618 CNRS, 1361, route des Lucioles, 
06560 Valbonne, France \\
$(c)$ Dipartimento di Fisica, Universit\`a di Firenze, and LENS, via Sansone 1, 50019 Sesto Fiorentino, Italy\\
$(d)$ Istituto Nazionale di Fisica Nucleare, Sezione di Firenze,  via Sansone 1, 50019 Sesto Fiorentino, Italy}

\begin{abstract}
The limits of applicability of the Lang-Kobayashi (LK) model for a semiconductor laser 
with optical feedback are analyzed. The model equations, equipped with realistic values 
of the parameters, are investigated below solitary laser threshold where 
Low Frequency Fluctuations (LFF) are usually observed. The numerical findings are compared with 
experimental data obtained for the selected polarization mode from a Vertical Cavity 
Surface Laser (VCSEL) subject to polarization selective external feedback. 
The comparison reveals the bounds within which the dynamics of the LK can be considered as realistic.
In particular, it clearly demonstrates that the deterministic LK, for realistic values of the linewidth 
enhancement factor $\alpha$, reproduces the LFF only as a transient dynamics towards one of the stationary 
modes with maximal gain. A reasonable reproduction of real data from VCSEL can be obtained only by considering 
noisy LK or alternatively deterministic LK for extremely high $\alpha$-values.
\end{abstract}
%
% explanation of PACS numbers:
% ----------------------------
%
%42.65.Sf Dynamics of nonlinear optical systems; optical instabilities,
%         optical chaos and complexity, and optical spatio-temporal dynamics
% 42.55.Px Semiconductor lasers; laser diodes
% 42.60.Mi Dynamical laser instabilities; noisy laser behavior
% 47.20.Ky Nonlinearity (including bifurcation theory)
% 05.45.Ra Coupled map lattices
% 05.45.-a Nonlinear dynamics and nonlinear dynamical
% systems (see also 45 Classical mechanics of
% discrete systems)
% 05.45.Pq Numerical simulations of chaotic models
% 68.10.Gw Interface activity, spreading
% 47.54.+r Pattern selection; pattern formation
% 02.30.Jr Partial differential equations
% 05.40.-a Fluctuation phenomena, random processes, noise, and Brownian motion

\maketitle 

%%%%%%%%%%%%%%%%%%%%%%%%%%%%%%%%%%%%%%%%%%%%%%%%%%%%%%%%%%%%%%%%%%%%%%
\section{Introduction}
\label{sec:0}
%%%%%%%%%%%%%%%%%%%%%%%%%%%%%%%%%%%%%%%%%%%%%%%%%%%%%%%%%%%%%%%%%%%%%%

The dynamics of semiconductor lasers with optical feedback is studied both 
experimentally and theoretically since almost 30 years (\cite{primolff},  for a review see e.g. \cite{book_lff} ). 
The interest for such configuration, commonly encountered in many applications (e.g, communication in optical fibers, 
optical data storage, sensing etc) arises from the rich phenomenology observed, ranging from multistability, 
bursting, intermittency, irregular and rare drops of the intensity (low frequency fluctuations (LFF)) 
and transition to developed chaos (coherence collapse (CC)). 
A complete understanding of the physical mechanisms at the basis of such complex 
behavior is, however, still lacking. In particular, the origin of the LFF regime is under 
debate since the very first observations and yet this puzzling problem has not been solved. 
Their origin was ascribed to stochastic effects~\cite{hk,hohl} or to deterministic but 
chaotic dynamics \cite{sano}, and more recently even to the interplay between regular periodic 
and quasi-periodic solutions \cite{david}. 
The LFF dynamics has been investigated by using several type of emitters, mainly edge-emitting: 
ranging from longitudinal multimode~\cite{multi_exp1,excit,multi_exp2,multi_teo1, multi_teo2, multi_teo3, solari} 
to single-mode DFB~\cite{dfb} semiconductor lasers. 

From the experimental point of view, a complete characterization of the LFF dynamics is quite difficult, 
because of the very different time-scales involved \cite{romanelli}. 
Indeed, fast oscillations on the 10-ps range have been observed 
with streak camera measurements \cite{streak1,streak2}, representing the fundamental scale on which 
the system evolves. On the other hand, the duration of such fast pulsing regime between LFF events can be 
as long as hundreds of nanoseconds or even microseconds. In literature, the LFF dynamics has been experimentally
characterized in several manners, starting from a relatively simple statistical analysis of the time 
separation T between LFF\cite{sukov,stat1,dfb} 
(relating the average $\langle T \rangle$ between LFF with the pump current) to 
Hurst exponents for the laser phase dynamics~\cite{lam}.

A widely used theoretical description of the system is the Lang-Kobayashi model \cite{LK}, introduced in
1980 in the effort to provide a simplified but effective analysis of an edge-emitting semiconductor 
laser optically coupled with a distant reflector. In the model, both the multiple reflections from the 
mirror (low coupling) and the possible multimodal structure of the laser were neglected. The opportunity to include 
such effects has been discussed in several papers 
\cite{balle_josab,solari,model_gen,multi_teo1,multi_teo2,multi_teo3}, 
but the model still remains presented 
as the standard theoretical approach to the system. 
While most of the phenomenology observed in the different experiments 
is grab by the model, quite often a more precise or quantitative comparison 
is obtained at the expense of a choice of parameters far from those actually measured or even not physically plausible.

Recently, a new configuration has been proposed and studied, based on a Vertical Cavity Surface Laser (VCSEL) with a 
polarized optical feedback \cite{romanelli}. Such a laser is longitudinal single-mode (due to the very short cavity) 
but may support different, high order transverse modes for strong enough pumping current 
(see e.g.\cite{book_vcsel}). The simmetry of the cavity allows 
also for the possible laser action on two different, linear polarizations selected by the crystal axis. 
The dynamics of the VCSEL with isotropical optical feedback has been examined experimentally  in
\cite{balle_josab, naumenko03} and theoretically in \cite{masoller}, while the role played
by polarized optical feedback has been discussed in \cite{besnard,loiko,loiko01}. In particular,
the setup used in \cite{romanelli} employed a polarizer in the feedback arm, in order to couple back only the radiation 
of one polarizations; moreover, a suitable range of pump current was chosen, to assure single transverse mode behavior. 
In such configuration, the appearance of LFF was reported and characterized. The possibility to control the role of the 
laser modes in the dynamics in this setup allows for a consistent description via the LK model and therefore
for an effective test of its predictions,  at variance with similar 
setup where instead the polarization selection were not used \cite{balle_josab,naumenko03}. 

Our aim in the present paper is to clarify the origin of the LFF dynamics by comparing numerical results obtained by 
solving the LK equations with experimental measurements done on a VCSEL. In particular, the parameters 
employed 
for the integration of the LK rate equations have been carefully derived from the analysis
 of the same VCSEL employed for the experiments~\cite{bar05}.

In Sec. II we describe our experimental setup, reporting the main phenomenology observed in the range 
of variation of the more relevant parameters of the system, namely, the pump current and the
phase of feedback.
The LK model is introduced and commented in Sec. III, together with the numerical methods 
employed for its integration and the choice of the parameter values derived from the experiment. 
In Sec. IV the properties of the stationary solutions are discussed, while in Sec. V 
a careful characterization of the deterministic model
is given, detailing the transient phenomena and the Lyapunov analysis. 
In Sec. VI the effect of noise is introduced and analyzed, discussing also its possible importance in the experiment.
A detailed comparison of the numerical results with the experimental measurements is given in Sec. VII, with particular 
regard to the distribution of the intensity and of the inter-event times for different parameter choice, 
including the alpha-factor and the acquisition bandwidths.
Finally, we draw our conclusions in Sec. VIII. 

%%%%%%%%%%%%%%%%%%%%%%%%%%%%%%%%%%%%%%%%%%%%%%%%%%%%%%%%%%%%%%%%%%%%%%
\section{Experimental Setup and Settings}
\label{sec:1}
%%%%%%%%%%%%%%%%%%%%%%%%%%%%%%%%%%%%%%%%%%%%%%%%%%%%%%%%%%%%%%%%%%%%%%

\begin{figure}[h]
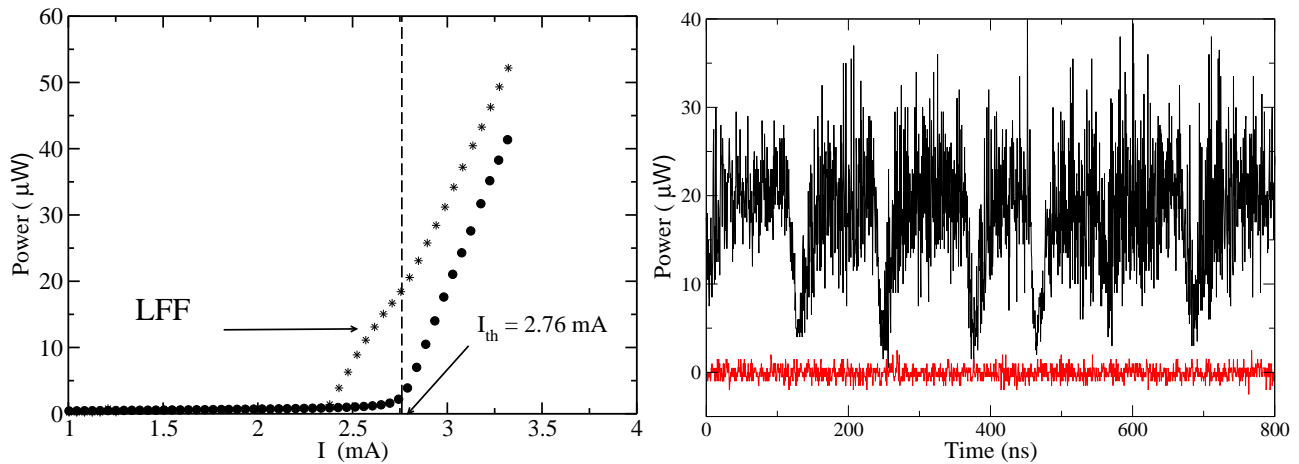

\centerline{
\includegraphics[draft=false, scale=.35,clip=true]{f1a}
\includegraphics[draft=false, scale=.35,clip=true]{f1b}
}
\caption{(a) Average power output versus the input current for the solitary laser (dots)
and for the laser with feedback (stars). (Color online) 
(b) Polarization modes: 
power outputs as a function of time for the VCSEL with feedback at
$I=2.64$~mA. Upper trace (black): main polarization; lower trace (red):
secondary polarization.
}
\label{fig:exp1}
\end{figure}

The experimental measurements are performed using
a VCSEL semiconductor laser with moderate polarized optical feedback. 
In particular, our analysis is limited to a regime
of pumping below the solitary threshold $I_{th} \sim 2.76$~mA, where 
the VCSEL emits light on a single linearly polarized transverse mode.
Longitudinal modes are not allowed by the cavity, and
other transverse modes are not present up to currents 
$\sim 6.5$~mA. The solitary laser emission
remains well polarized up to roughly the same current (see Fig. \ref{fig:exp1} (b)).

The technical details of the source are the following. The laser is
an air-post VCSEL made by CSEM \cite{gulden}, operating around 770 nm.
The mesa diameter is 9.4 $\mu$m, a ring contact defines the ouput
window with a diameter of 5 $\mu$m, and the active medium is composed
of three 8 nm quantum wells. The temperature of the laser case is stabilized
within 1 mK, the pump current is controlled by a home made battery operated
power supply whose current noise is below 40 pA/Hz$^{-1/2}$ in the frequency
range from 1kHz to 3 MHz. 

The feedback is applied on the polarization direction of the solitary laser emission. 
The external cavity includes collimation optics, two polarizers, a variable attenuator, 
and the feedback mirrors that is mounted on a piezoelectric transducer at about 50~cm 
from the laser. The output radiation, after optical isolators, is detected by an avalanche 
photodiode with a bandwidth of about 2~GHz, whose signal, sometimes after low-pass filtering, 
is recorded by a 4~GHz bandwidth digital scope.
More details on the experiment can be found in Refs~\cite{romanelli,Soriano2004}. 

Optical feedback results in a reduced threshold $I_{th}^{red} \sim 2.42$~mA, as shown in Fig.~\ref{fig:exp1}~(a). 
We have examined the dynamic behavior of the output intensity for various pump currents, 
both above and below $I_{th}$, with a particular attention to possible effects of the feedback 
phase $\Delta\phi$ which is varied by acting on the external mirror piezoelectric transducer.

\begin{figure}[h]
\includegraphics[draft=false,scale=.52,clip=true]{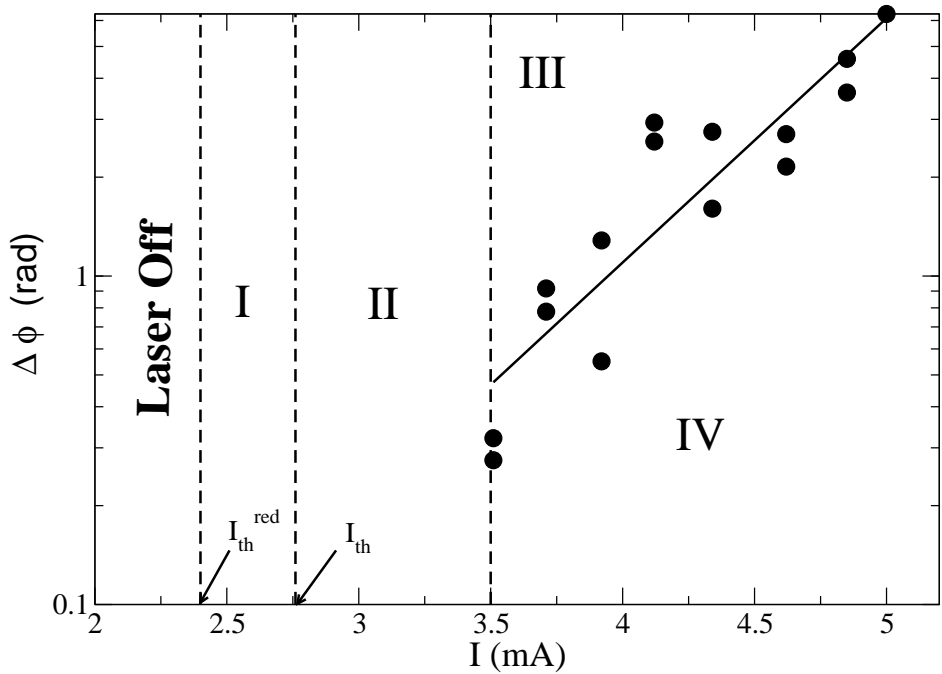}
\caption{Phase diagram of the VCSEL with feedback: phase of
the feedback $\Delta \phi$ as a function of the pump current $I$. 
The roman numbers denote regions of different dynamical regimes:
(I) single-mode LFF, (II) two-mode LFF, (III) coherence collapse
and (IV) stable emission. The vertical dashed line indicates (from
low to high current) successive current thresholds:
the reduced one $I^{red}_{th}$, the solitary laser one $I_{th}$ and 
the current value separating LFF from coherence collapse. 
The dots represent the maximal $\Delta \phi$ for which the laser emission
remains stable, while the solid line
is a guide for eyes to distinguish regions (III) from (IV).}
\label{fig:exp2}
\end{figure}

As shown in Fig. \ref{fig:exp2},
we can identify several regimes: (I) at $I < I_{th}$ one observes
a single mode LFF dynamics (i.e. the main polarization exhibits LFFs,
while the secondary polarization remains off,
see Fig. \ref{fig:exp1} (b)); (II) $ I_{th} < I < 3.5$~mA: in this regime the
LFF dynamics of the main polarization is accompanied by a synchronized spiking behavior 
in the secondary polarization (coupled modes LFF). This regime was analyzed in Ref.~\cite{romanelli} and, 
in more details, in Ref.~\cite{Soriano2004}, where it is proposed that the dynamics of the 
secondary polarization is driven by the main polarization, whose behavior is not influenced by 
the orthogonal polarization mode.
For larger pump current one begins to observe Coherence Collapse (regime (III) in Fig. \ref{fig:exp2})
and moreover the feedback phase begins to play 
a fundamental role. In particular, for increasing $I$ in larger and larger portion 
of the phase interval the laser is stationary
(regime (IV) in Fig. \ref{fig:exp2}). This phenomenon
is not yet understood and it will be subject of a future analysis \cite{loiko}.

Anyway, in the present paper we will limit our analysis
to regime (I) (i.e. for $I < I_{th}$), where the VCSEL has a single-mode LFF 
dynamics and the phase delay of the feedback does not play any role.
We remark that this statement is suggested by the experiment, where
the phase stability would be enough to discriminate such effects,
as shown in the analysis of regimes (III-IV).

%%%%%%%%%%%%%%%%%%%%%%%%%%%%%%%%%%%%%%%%%%%%%%%%%%%%%%%%%%%%%%%%%%%%%%
\section{Numerical model and methods}
\label{sec:2}
%%%%%%%%%%%%%%%%%%%%%%%%%%%%%%%%%%%%%%%%%%%%%%%%%%%%%%%%%%%%%%%%%%%%%%

The dynamics of the VCSEL for $I < I_{th}$ is a purely single mode dynamics
and therefore we expect that it could be reproduced by employing the  Lang-Kobayashi (LK)~\cite{LK}
rate equations for the complex field $E(t)$ and the carrier density $n(t)$. 
In order to achieve an accurate and reliable comparison of the numerical results 
with the experimental ones we will employ for our simulations the laser 
working parameters reported in Table~\ref{parameter}. These parameters have been
determined via a series of suitable experiments for exactly the same VCSEL 
employed to obtain the measurements examine in this paper~\cite{bar05}.
The only lacking parameter that was not possible to obtain in the previous characterization 
is the feedback strength, which is however determined from the threshold reduction. 

In this article we use the rate equations derived in Ref.~\cite{bar05}
for the single-mode solitary laser, modified to include the feedback. Defining  
the deviation $\Delta n$ of the carrier
density from transparency normalized to have unitary value at threshold,
i.e.
\begin{equation}
\Delta n = \frac{n-1}{n_{th}-1}
\qquad ,
\label{carrier}
\end{equation}
the following form is obtained~\cite{bar05}:
\begin{eqnarray}
\tau_n \dot {\Delta n} &=& - \Delta n + 1 + \eta(\mu-1) -\Delta n|E|^2 
\nonumber
\\
\dot E &=& \frac{1 + i \alpha}{2 \tau_p} [\Delta n-1]E
+ \frac{k}{\tau_p} {\rm e}^{-i \omega \tau} E(t - \tau) +\sqrt{R_0} \tilde \xi(t)
\label{model}
\end{eqnarray}
where $\tilde \xi(t)= \xi_R(t) + i \xi_I(t)$ is a complex Gaussian noise term with zero mean 
and correlation given by $<\xi_R(t) \xi_R(0)> = <\xi_I(t) \xi_I(0)>= \delta(t)$
and $<\xi_R(t) \xi_I(0)>= 0$.  The noise variance $R_0=(n/n_{th})^2 R_{sp}$ 
represents a multiplicative noise term proportional to the square of the reduced
carrier density $n/n_{th}$ ($n_{th}$ being the threshold carrier density) and
to the variance of the spontaneous emission noise $R_{sp}$.
The parameter $\mu=I/I_{th}$ is the pump current rescaled to unity
at threshold, $\tau_n$ and $\tau_p$ are carrier and photon lifetimes, respectively,
$\tau$ is the delay (or external roundtrip time), $\alpha$ is the 
linewidth enhancement factor and $\eta$ the reduced gain (for the exact definitions
of these quantities in terms of the laser parameters
and for the approximations employed to derive (\ref{model}) see Ref.~\cite{bar05}).

By reexpressing the time scale in terms of the photon life-time ($\tau_p$)
the equations assume the usual form for the LK model and read as
\begin{eqnarray}
T  {\dot \Delta n} &=& -\Delta  n + p -\Delta n |E|^2 
\nonumber
\\
\dot E &=& \frac{1 + i \alpha}{2} [\Delta n -1] E
+ k {\rm e}^{-i \omega \tau} E(t - \tau)
+\sqrt{R} \tilde \xi(t)
\label{model1}
\end{eqnarray}
where $T=\tau_n / \tau_p$, $p=1+\eta (\mu -1)$ and
$R = (n/n_{th})^2 R_{sp} \times \tau_p$.

The complex field can be expressed as 
$E(t) = A(t) +i B(t) = \rho(t) \exp{i \psi(t)}$
and the equations can be rewritten as:
\begin{eqnarray}
\dot A(t) &=& \left(\frac{\Delta n(t) -1}{2}\right) \left(A(t) +\alpha B(t) \right)
+ k [A(t -\tau) \cos{\phi} +B(t-\tau)\sin{\phi}]
+\sqrt{R/2} \xi_R(t) 
\nonumber
\\
\dot B(t) &=& \left(\frac{\Delta n(t) -1}{2} \right)\left(B(t) - \alpha A(t) \right)
+ k [B(t-\tau)\cos{\phi} -A(t -\tau) \sin{\phi} ]
+\sqrt{R/2} \xi_I(t) 
\nonumber
\\
T  {\Delta \dot n}(t)&=& p -\Delta  n (t) \left[1+A^2(t)+B^2(t) \right]
\label{model2}
\end{eqnarray}
where $\phi= \omega \tau$.

Moreover, by assuming that $n \sim n_{th}$ the noise terms become additive 
with an adimensional variance $R = 2.76 \times 10^{-3}$, 
the other quantities entering in (\ref{model2}), expressed  in $\tau_p$ units, are
$\tau = 302.5$, $ \omega \times \tau=8.743 \times 10^6$ and  $T = 30.8333$,
the numerical values have been obtained by
employing the parameters values in Table ~\ref{parameter}.

In order to reproduce the power-current response curve for two different experimental 
data sets we have chosen feedback strengths $k =0.25$ and 0.35, 
while typically we considered pump currents and linewidth enhancement factors 
in the range $0.9 \le \mu  \le 1.20$,  and $3 \le \alpha \le 5$, respectively.

 The deterministic equations (\ref{model2}) with $R\equiv0$ 
have been integrated by employing the method introduced by Farmer
in 1982 \cite{farmer} equipped with a standard 4th order Runge-Kutta
scheme, while for integrating the equations with the stochastic terms
we have employed an Heun integration scheme~\cite{toral}.
The simulations have been performed by integrating the field
variables with time steps of duration $\Delta t = \tau/(N-1)$,
with $N =1,000 - 10,000$.

The dynamical properties of the system can be
estimated in terms of the associated Lyapunov spectrum, that fully 
characterizes the linear instabilities of infinitesimal perturbations 
of the reference system. By following the approach reported in \cite{farmer},
we have estimated the Lyapunov spectrum $\{ \lambda_k \} \quad (k=1,\dots,2N+1)$ 
by integrating the linearized dynamics associated to eqs (\ref{model2}) 
in the tangent space and by performing periodic Gram-Schmidt 
ortho-normalizations according to the method reported in~\cite{benettin}.
The Lyapunov eigenvalues $\lambda_k$ are real numbers 
ordered from the the largest to the smallest, a positive maximal Lyapunov $\lambda_1$
is a indication that the dynamics of the system is chaotic.
Moreover, from the knowledge of the Lyapunov spectrum it is possible 
to obtain an estimation of the number of degrees of freedom actively involved 
in the chaotic dynamics in terms of the Kaplan-Yorke dimension~\cite{ky}:
\begin{equation}
D_{KY} = j + \frac{\sum_{k=1}^j \lambda_k}{|\lambda_{j+1}|}
\label{d_ky}
\end{equation}
where $j$ is the maximal index for which $\sum_{k=1}^j \lambda_k \ge 0$.

\begin{table}[ht]
\caption[tabone]{
Experimental values of the parameters entering in the 
model (\ref{model}) (from \protect\cite{bar05}).
}
%\vskip 0.3 truecm
\begin{ruledtabular}
\begin{tabular}{lrr}
Description \hfil & \hfil Symbol \hfil & \hfil Value \hfil \\
\hline
Linewidth enhancement factor &   $\alpha$ & $3.2 \pm 0.1$   \\
Photon lifetime in the cavity&   $\tau_p$ & $12 \pm 1 \enskip ps$   \\
Carrier lifetime&   $\tau_n$ & $0.37 \pm 0.02 \enskip ns$   \\
External roundtrip time&   $\tau$ & $3.63 \enskip ns$   \\
Variance of the spontaneous emission noise 
&$R_{sp}$ & $(2.3 \pm 0.6) \times 10^{-4} \enskip ps^{-1}$   \\
Reduced gain &   $\eta$ & $5.8 \pm 0.6$   \\
\end{tabular}
\end{ruledtabular}
\label{parameter}
\end{table}

%%%%%%%%%%%%%%%%%%%%%%%%%%%%%%%%%%%%%%%%%%%%%%%%%%%%%%%%%%%%%%%%%%%%%%
\section{Stationary Solutions}
\label{sec:3}
%%%%%%%%%%%%%%%%%%%%%%%%%%%%%%%%%%%%%%%%%%%%%%%%%%%%%%%%%%%%%%%%%%%%%%

A first characterization of the phase space of the LK system can be achieved
by individuating the corresponding stationary solutions and by analyzing
their stability properties. The stationary solutions of the above set of 
equations can be found by
setting $\dot \rho = \Delta \dot n = 0$ and
${\dot \psi} = \Omega$,
i.e. by looking for solutions of the form
\begin{equation}
E_S(t) = \rho_S {\rm e}^{i \Omega t} 
\qquad {\rm and} \qquad \Delta n(t) = \Delta n_S \quad .
\label{staz}
\end{equation}
The stationary solutions can be parametrized in terms of
the variable $\theta= \phi + \Omega \tau$, and by setting $J = (p-1)/2$
they can be expressed as follows \cite{yanchuk}
\begin{equation}
X_S = \frac{(\Delta n_S-1)}{2} = - k \cos(\theta)
\end{equation}
\begin{equation}
\rho_S^2  = 2 \frac{J-X_S}{2 X_S +1} \ge 0
\end{equation}
and
\begin{equation}
\Omega = \frac{\theta -\phi}{\tau} = - k
\sqrt{1+\alpha^2} \sin(\theta + \theta_0)
\end{equation}
where $\theta_0 = {\rm atan}({\alpha})$ and $-\pi < \theta \le\pi$.

The  stationary solutions are usually termed external cavity modes (ECMs) and
correspond to stationary lasing states, in the limit of long delay they 
are densely covering the following ellipse:
\begin{equation}
k^2 = X_S^2 +(\Omega -\alpha X_S)^2
\quad .
\label{ellisse}
\end{equation}

The stability properties of these solutions can be obtained by estimating
the associated Floquet spectrum  $\{\Lambda_n\}$, these eigenvalues are typically
complex and their number is infinite, due to the delay term present in the
LK equations. However, the stability properties of the ECMs are determined
by the eigenvalues with the largest real part, in particular by the
maximal one $\Lambda^M = ({\rm Re} \enskip \Lambda^M, {\rm Im} \enskip \Lambda^M)$.
These can be easily determined by solving the characteristic equation obtained by 
linearizing around a certain ECM:
\begin{eqnarray}
& &Z^2_n [\Lambda_n + \varepsilon(1+\rho_S^2)] +2 Z_n [\Lambda_n \cos(\theta)( \Lambda_n+\varepsilon(1+\rho_S^2) 
+ \varepsilon \rho_S^2(1- 2k \cos(\theta))(\cos(\theta)-\alpha \sin(\theta))/2] 
\nonumber \\
&+& \Lambda_n^2[\Lambda_n +\varepsilon(1+\rho_S^2)] + \Lambda_n \varepsilon \rho_S^2 (1-2k \cos (\theta))
\equiv a_n Z^2_n + 2b_n Z_n + c_n = 0
\quad .
\label{charact}
\end{eqnarray}
where $Z_n = k (1 -{\rm e}^{- \Lambda_n \tau})$ and $\varepsilon = 1/T$.
The {\it pseudo-continous} spectrum~\cite{ya2} can be obtained by solving eq. (\ref{charact}) in terms of $Z_n$ and
by considering $\Lambda_n$ as a parameter.
In this case the solutions of the corresponding second order equation are
\begin{equation}
\Lambda_n^{\pm} = - \frac{1}{\tau} \log\left[  1 + \frac{b_n \mp \sqrt{b_n^2 -a_n c_n}}{k a_n} \right ] + i 2 \pi n
\label{sol}
\end{equation}
and by self-consistently solve the above equation one can find all the eigenvalues, that are arrangend in
two branches.  However, the spectrum contains also isolated eigenvalues, that can be obtained by
solving directly Eq. (\ref{charact}) in terms of $\Lambda_n$ and by
considering $Z_n$ as a parameter. In this case the equation is
a cubic and by solving self-consistely its expression one finds (up to) three distinct eigenvalues. 
While the pseudo-continuos spectrum emerges due to the presence of the delay in the system,
the isolated eigenvalues originate from those characterizing the three dimensional single 
laser rate equations in absence of the delay, i.e., the Eq. (\ref{model1}) with $\tau=0$~\cite{ya2}.

\begin{figure}[h]
\centerline{
\includegraphics[draft=false, scale=.35,clip=true]{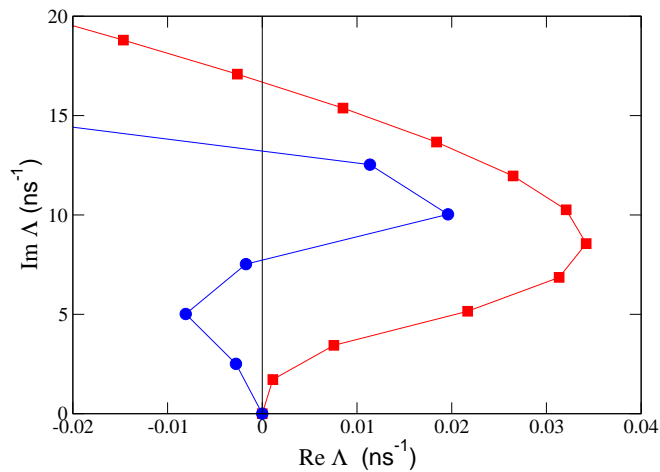}
}
\caption{(Color online) Unstable branch $\Lambda$ associated to specific modes
of the LK. The mode $\theta=-0.32$ for $\alpha=3.20$ reveals a sort of
modulational instability (red squares), while the mode $\theta=-0.08$ for $\alpha=3.22$
shows a Turing-like instability (blue circles). The results refer to  $\mu=0.93$. Due
to the symmetry of the spectra only the part corresponding to $Im \Lambda > 0$ is displayed.
}
\label{fig:spectra}
\end{figure}

Typically, depending on their linear stability properties ECM are divided into 
{\it modes} and {\it antimodes} \cite{levine}. Antimodes
are characterized by a positive real eigenvalue and are therefore
unstable. For the modes instead the maximal real eigenvalue is zero and they
are unstable whenever the associated spectrum crosses the imaginary axis.
Various types of instabilities can be observed for these delayed systems and 
they can be classified for analogy with
the spatially extended systems as for example {\it modulational-type} or 
{\it Turing-type} instabilities~\cite{yanchuk,ya2}.
Examples of the unstable branch for the spectra $\{\Lambda_k\}$ associated to Turing-type 
and modulationali-type instabilities are reported in Fig. \ref{fig:spectra}, for a classification
of the possible instabilities of equilibria for delay-differential equations see \cite{ya2}.

In particular, the modes correspond to $\theta$-values located within the interval
$[\theta_2 : \theta_1]$, where $\theta_1 \sim {\rm atan} (1/\alpha)$ and
$\theta_2 \sim -\pi +{\rm atan} (1/\alpha)$. 
Moreover, stationary solutions are acceptable only 
if they correspond to positive intensities $\rho^2_S \ge 0$, i.e. if they 
are situated within the interval $[-\theta_R:\theta_R]$ with $\theta_R = 
{\rm acos}{(-J/k)}$.
In the range of parameters examined in the present paper the number
of stable modes (SMs) is always between two and four and they are
located in a narrow interval of $\theta$ located around zero (i.e.
around the so called Maximum Gain Mode (MGM)).

\begin{figure}[h]
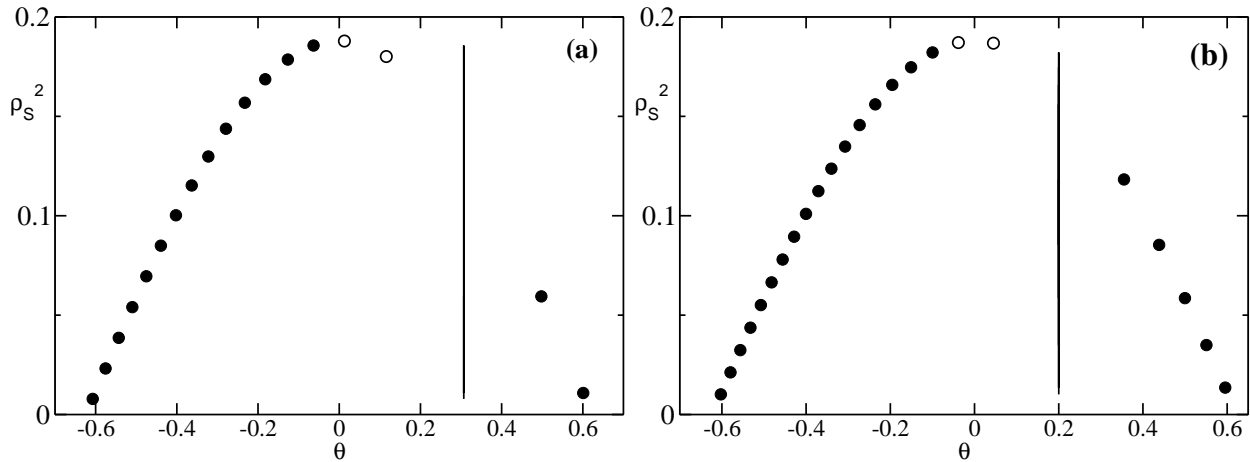

\centerline{
\includegraphics[draft=false, scale=.35,clip=true]{f4a.eps}
\includegraphics[draft=false, scale=.35,clip=true]{f4b.eps}
}
\caption{Intensities versus $\theta$ for
ECMs of eq. \protect(\ref{model2}) with $\mu=0.93$ and 
for $\alpha=3.2$ (a) and $5.0$ (b).
The solid line indicates $\theta_1$.
The ECMs with $\theta < \theta_1$ are modes, while the others are
antimodes. The empty circles indicates the SMs towards
which we numerically observed relaxation of the system by considering
up to 100 different random initial conditions.
}
\label{fig:modes}
\end{figure}

As we will report in the following section, by integrating the 
deterministic version of the model (\ref{model2}) for moderate
$\alpha$- and $\mu$-values, we always observe a relaxation towards one of 
the SMs. In particular for $3 < \alpha < 5$ the dynamics
seem to relax always towards one of the SMs located 
in proximity of the MGM (corresponding to $\theta \equiv 0$) 
(see Fig. \ref{fig:modes}). It should be noticed that the MGM is
a solution of the system only for appropriate choices of 
$\phi$. Two examples of typical ECMs are reported in Fig. \ref{fig:modes}
for the present system for $\alpha=3.2$ and 5 and $\mu=0.93$, in both
cases two stable attracting modes have been identified.
Recently, a similar coexistence of two stable solutions located in proximity of the MGM
has been reported experimentally for an edge-emitter laser with a low level of optical 
feedback~\cite{mendez}

As reported in \cite{yanchuk} the stability properties of the 
MGM do not depend on $\alpha$, therefore it is reasonable
to expect that some SM will be always present in a narrow window
around $\theta$ for any chosen linewidth enhancement factor~\cite{levine} 
and that they will coexist with the chaotic dynamics, as observed 
experimentally in \cite{heil}.

%%%%%%%%%%%%%%%%%%%%%%%%%%%%%%%%%%%%%%%%%%%%%%%%%%%%%%%%%%%%%%%%%%%%%%
\section{Deterministic Dynamics}
\label{sec:4}
%%%%%%%%%%%%%%%%%%%%%%%%%%%%%%%%%%%%%%%%%%%%%%%%%%%%%%%%%%%%%%%%%%%%%%

  An open problem concerning the deterministic LK equations is if and for which range
  of parameters these equations faithfully reproduce the experimentally 
  observed dynamics. Particular interest is usually focused on the 
  $\alpha$-value to be employed to obtain a realistic behaviour of the
  model.

%%%%%%%%%%%%%%%%%%%%%%%%%%%%%%%%%%%%%%%%%%%%%%%%%%%%%%%%%%%%%%%%%%%%%%
\subsection{LFFs as a transient phenomenon}
\label{sec:4.1}
%%%%%%%%%%%%%%%%%%%%%%%%%%%%%%%%%%%%%%%%%%%%%%%%%%%%%%%%%%%%%%%%%%%%%%

\begin{figure}[h]
\includegraphics[draft=false, scale=.42,clip=true]{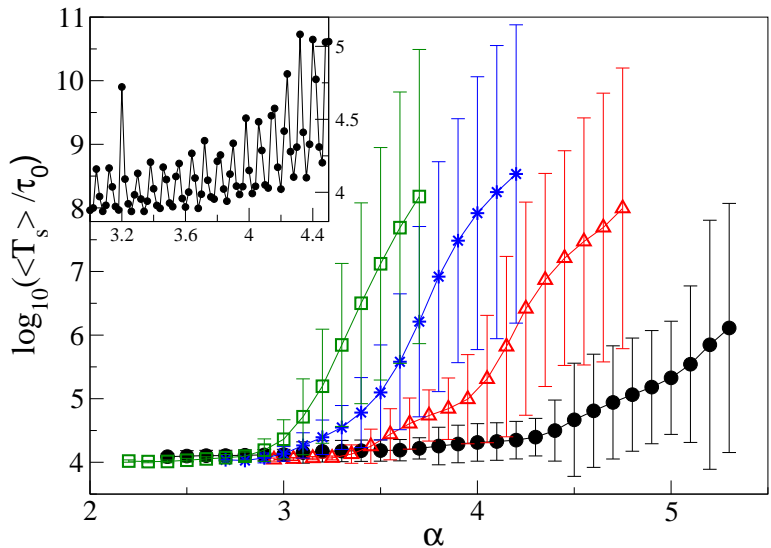}
\caption{(Color online) 
Logarithm of the transient times $<T_s>$ as a function
of $\alpha$ for various $\mu$-values:  namely, $\mu=0.93$
(black filled circles), $0.95$ (red empty triangles),  $0.97$ 
(blue asteriskes)
and $0.99$ (green empty squares). The data refers to $k=0.25$,
$\Gamma=10^{-5}$, $t_w = 1000 \tau = 3.630 \mu s$ and $M \sim 10-100$, 
and they have been obtained by employing an integration time step 
$\Delta t=3.63 ps$.
The time scale of the figure is $\tau_0 = 1 \enskip$ ns.
The bars reported for each measured value indicate the
range of variability of $<T_s>$ (within a $\alpha$-interval
of amplitude 0.1) due to its finer structure shown in the inset.
In the inset are displayed the transient times for $\mu=0.93$
reported at a higher resolution in $\alpha$, namely for a
resolution of $0.02$. 
}
\label{fig:times}
\end{figure}

  To answer this question we consider the deterministic verson of the 
  model (\ref{model2}) (i.e. by assuming $R\equiv 0$)
  equipped with the parameter values deduced by the experimental
  data \cite{bar05} apart for $\mu$ and $\alpha$ that will be varied.
  In particular, we would like to understand if the system
  shows a LFF behaviour and if such dynamics is statistically stationary
  or not. In order to verify it, we initialize randomly 
  the amplitude $\rho(t=0)$, the phase $\phi(t=0)$ and
  the excess carrier density $\Delta n(t=0)$, then we follow
  the dynamics and we examine the evolution of the field intensity
  $\rho^2(t)$. If the dynamics end up on a stationary solution
  we register the time $T_s$ necessary to reach it and we average this
  time over many ($M$) different initial conditions.
  In order to measure $T_s$, we estimate the time needed to the 
  standard deviation of the intensity 
  (evaluated over sub-windows of time duration $t_w$)
  to decrease below a chosen threshold $\Gamma$.
  The average times $<T_s>$ are reported in Fig. \ref{fig:times}

  The main result shown in Fig.~\ref{fig:times} is that the 
system after a transitory phase (shorter or longer) settles down to
a SM and that the duration of the transient increases for
increasing $\alpha$ or $\mu$-values. Preliminary indications
in this direction have been previously reported in \cite{dfb}.
These results clearly indicate that the LFF dynamics is just a transient phenomenon
for the deterministic LK equation for commonly employed $\alpha$-values
(i.e., for $\alpha \sim 3.0-3.5$).
For each fixed $\mu$
it seems that $<T_s>$ diverge above some "critical" $\alpha$-value,
however we cannot exclude that the system will also in this case
finally converge to a SM. At least we can consider
the reported critical values as lower bounds above which LFF
could occur as a stationary phenomenon and not as a transitory state.
With the chosen numerical accuracy and due to the available
computational resources, for any pratical purpose a transient longer 
than $0.1 -1$ s can be considered as an infinite time.

\begin{figure}[h]
\includegraphics[draft=false, scale=.35,clip=true]{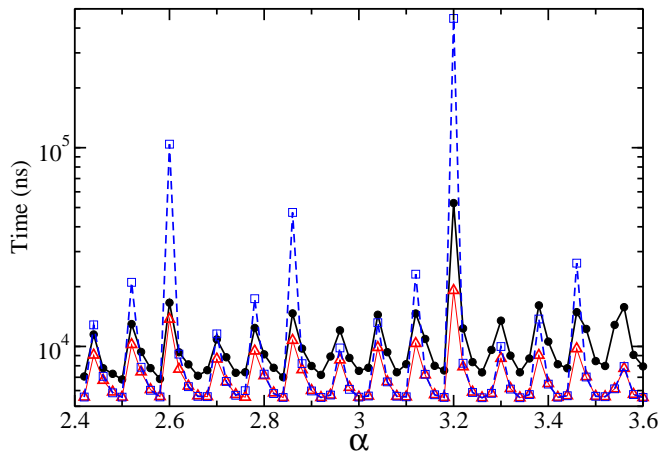}
\caption{(Color online) 
Transient times $<T_s>$ (black filled circles), 
$T_{est}$ (red empty triangles) and $T_{1}$ (blue empty square)
in nanoseconds as a function of $\alpha$.
The results have been obtained by examining the decay of $\rho^2(t)$ 
with $\Gamma=10^{-5}$ and $t_w = 1000  \tau = 3.630  \mu s$ 
and $M \sim 100-500$. The employed
integration time step is $\Delta t=0.363$ ps and the data
refer to $k=0.25$ and $\mu=0.93$.
}
\label{fig:times_stablemodes}
\end{figure}

 An interesting feature is that the variability of $<T_s>$
with $\alpha$ is quite wild and it reflects the stability of
the modes in proximity of the MGM, as shown in Fig. \ref{fig:times_stablemodes}.
However the average values $<T_s>$ (averaged also 
over $\alpha$-intervals of width $0.1$) still indicates a clear trend 
of the transient times to increase with $\alpha$.
 
We expect that the time scale associated to the convergence
to a stable mode is ruled by the eigenvalue with the maximal
real part (apart the eigenvalue zero, that is always present
due to the phase invariance of the LK equations). In particular
we expect that the intensity of the signal will converge towards the
stable mode as
$$
\rho^2(t) \sim \rho^2(0) {\rm e}^{\rm 2 Re \enskip \Lambda^M \enskip t} cos(\rm  2 Im \enskip \Lambda^M \enskip t)
+ \rho_S^2
$$
where $\Lambda^M$ is the eigenvalue with maximal (non zero) real part 
associated to the considered stable solution.
For small $\mu$ we have observed that
the dynamics can collapse to different stable solutions (typically,
from 1 to 3). By estimating the probability $P_m=N_m/M$ to end up in one of these states
(being $N_m$ the number of initial conditions converging to the $m$-mode)
and by indicating the corresponding eigenvalue with maximal real part as $\Lambda^M (m)$, a reasonable
estimate of $<T_s>$ is given by
\begin{equation}
T_{est} = \frac{\ln{\Gamma}}{2} \sum_m \frac{P_m}
{|\rm Re \Lambda^M (m)|}  \quad ,
\label{test}
\end{equation}
where $\Gamma$ is the employed threshold.
As it can be seen in Fig. \ref{fig:times_stablemodes}, the estimation
is quite good and the periodicity of the
two quantities is identical in the examined range of $\alpha$-values
and for $\mu = 0.93$. The expression (\ref{test}) always gives
a good estimate of $<T_s>$  for $\alpha <4$ and for $\mu < 1$,
but the agreement worsens for increasing $\alpha$-values.

However, even a more rough estimate is capable
to give a reasonable approximation of $<T_s>$
and in particular to capture its periodicity.
This estimate is simply given by
\begin{equation}
T_{1} = \frac{\ln{\Gamma}}{2} \frac{1}
{|\rm Re \bar \Lambda^M|}  \quad ,
\label{max}
\end{equation}
where $\bar \Lambda^M$ is the eigenvalue with maximal non zero real part associated
to the stable mode with higher $\theta$ (i.e. the first stable
mode located in proximity of the anti-mode boundary).

From the present analysis it emerges that the stable modes play
a relevant role for the (transient) dynamics of the deterministic LK equations
at least for $\alpha < 4$. Moreover the observed strongly fluctuating
behaviour of $<T_s>$ as a function of $\alpha$ indicates that 
the choice of this parameter is quite critical, since a small variation can 
lead to an increase of an
order of magnitude of the transient time\footnote{For $\mu=0.97$
a variation of $\alpha$ from 3.48 to 3.60 leads to an increase
of $<T_s>$ from 130 $\mu$s to 1.3 ms.}

%%%%%%%%%%%%%%%%%%%%%%%%%%%%%%%%%%%%%%%%%%%%%%%%%%%%%%%%%%%%%%%%%%%%%%
\subsection{Lyapunov Analysis}
\label{sec:4.2}
%%%%%%%%%%%%%%%%%%%%%%%%%%%%%%%%%%%%%%%%%%%%%%%%%%%%%%%%%%%%%%%%%%%%%%

\begin{figure}[h]
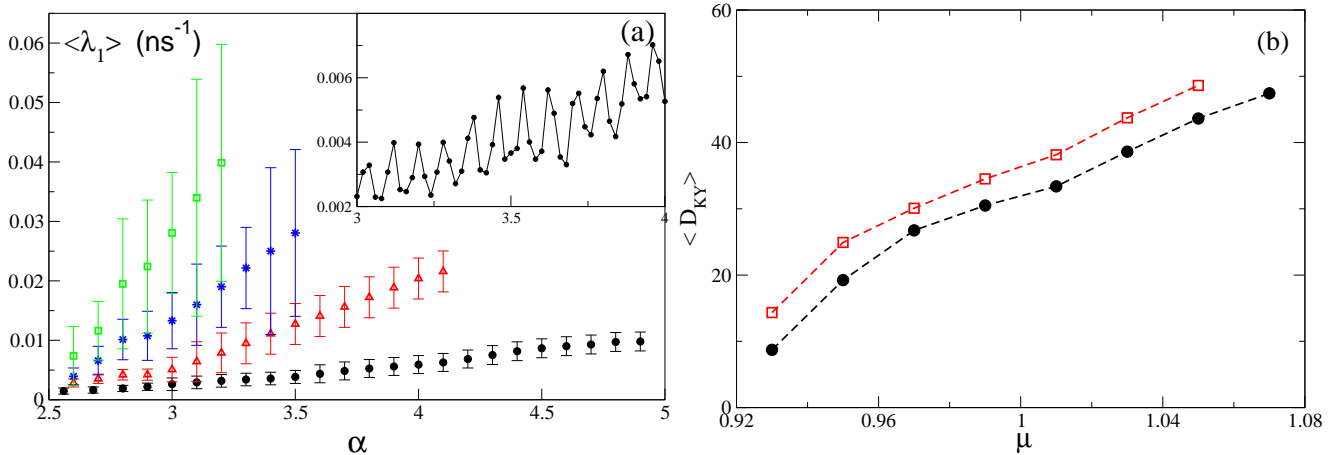

\centerline{
\includegraphics[draft=false, scale=.35,clip=true]{f7a.eps}
\includegraphics[draft=false, scale=.35,clip=true]{f7b.eps}
}
\caption{(Color online) (a) Average maximal Lyapunov exponents $<\lambda_1>$ as a function
of $\alpha$ and for various $\mu$-values below threshold: 
namely, $\mu=0.93$
(black filled circles), $0.95$ (red empty triangles),  $0.97$ (blue asterisks)
and $0.99$ (green empty squares). 
The bars reported for each measured value indicates the
range of variability of $<\lambda_1>$ due to its finer structure
as measured within a $\alpha$-interval of width 0.1.
In the inset the data for $<\lambda_1>$ are reported
for a higher resolution in $\alpha$ (namely 0.02) for $\mu=0.93$.
(b) Average Kaplan-Yorke dimensions
$<D_{KY}>$ as a function of $\mu$ for $\alpha=4$ (black filled circles) 
and 5 (red empty squares).
All the data refer to $k=0.25$, for the $<\lambda_1>$ estimation
$\Delta t=0.363$ ps, $M=500$, and $T_{int}=3.63$ ms while for the $<D_{KY}>$ evaluation
$\Delta t=3.63$ ps, $M=20$ and $T_{int}=0.14$ ms.
}
\label{fig:lyap}
\end{figure}

We have characterized the transient dynamics preceeding the
collapse in the stationary state in terms of the maximal
Lyapunov $\lambda_1$ and of the associated Kaplan-Yorke (or Lyapunov)
dimension $D_{KY}$. In particular, these quantities reported in Fig. \ref{fig:lyap}
have been estimated by integrating the linearized dynamics for a sufficiently long 
time period $T_{int}$ and by averaging over $M$
different initial realizations.

 It is clear from the figures that the average maximal Lyapunov
 exponent $<\lambda_1>$ is definetely not zero for all the considered
 situations and that it increases (almost steadily) with the parameter
 $\alpha$ as well as with the pump parameter. Moreover, also this indicator
 reflects the stability properties of the SMs located in proximity
 of the MGM, by exhibiting large oscillations as a function of $\alpha$,
as shown in the inset of \ref{fig:lyap}(a).

 The values of $<D_{KY}>$ reported in \ref{fig:lyap}(b)
 clearly indicate that the system cannot be described as low
 dimensional, even during the transient and 
 even below the solitary laser threshold. As a matter of fact
 the number of active degrees of freedom
 ranges between $10$ and $50$. It should be noticed that
 $<D_{KY}>$ is determined by the instability properties 
 not only of anti-modes and but also of modes that
 have bifurcated (via Hopf instabilities) becoming unstable for increasing $\alpha$.

%%%%%%%%%%%%%%%%%%%%%%%%%%%%%%%%%%%%%%%%%%%%%%%%%%%%%%%%%%%%%%%%%%%%%%
\section{Noisy Dynamics}
\label{sec:5}
%%%%%%%%%%%%%%%%%%%%%%%%%%%%%%%%%%%%%%%%%%%%%%%%%%%%%%%%%%%%%%%%%%%%%%

\begin{figure}[h]
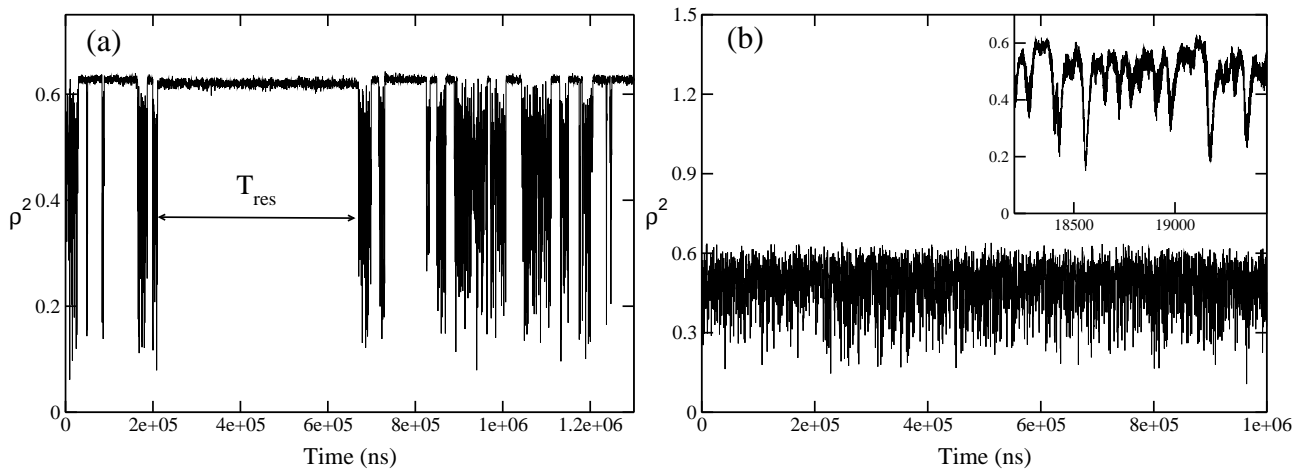

\centerline{
\includegraphics[draft=false, scale=.35,clip=true]{f8a.eps}
\includegraphics[draft=false, scale=.35,clip=true]{f8b.eps}
}
\caption{Intensity of the field $\rho^2$ as a function of the time
for the noisy LK equations. The data have been filtered 
with a low-pass filter at $80$~MHz and refer to $\mu=0.97$, $\alpha=3.3$.
The results reported in (a) correspond to a noise variance
$R=3 \times 10^{-4}$, while those in (b) are relative to
$R=3 \times 10^{-3}$. In (a) a typical time of residence $T_{res}$
around one of the SMs is indicated. The inset in (b)
is an enlargement of the actual dynamics.
}
\label{fig:interm}
\end{figure}

\begin{figure}[h]
\includegraphics[draft=false, scale=.45,clip=true]{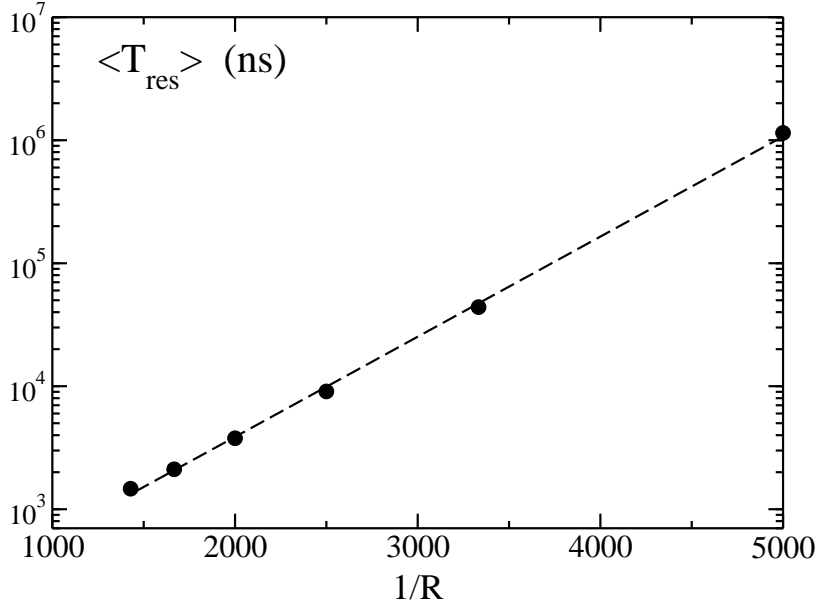}
\caption{Average residence times in the SM as a function of the
inverse of the variance of the additive noise to the LK equations.
The dashed line is a exponential fit $\propto \exp{[W/R]}$ to the numerical data, the
fitted exponential slope is $W=0.00187$.  The data refer to $\mu=0.97$ and
$\alpha=3.3$.
}
\label{fig:escape}
\end{figure}

 We have examined the
 dynamics (\ref{model}) for increasing level of noise, namely
 for $10^{-6} < R < 10^{-2}$. Also in this case, for $\alpha=3.3$ and
 for noise levels smaller than $10^{-3}$ the LFF dynamics only occurs during 
 a transient. However for increasing $R$ values we observed
 a transition to sustained LFF and the transition region was
 characterized by an intermittent behaviour. These behaviours
 are exemplified in Fig. \ref{fig:interm} for $\alpha=3.3$
 and $\mu=0.97$.  As shown in Fig. \ref{fig:interm}(a) the
 orbit spends long times in proximity of one of the SM and then,
 due to noise fluctuations, escapes from the attraction basin
 associated to the stable solution and exhibit LFFs before being
 newly reattracted by the SM. This intermittent dynamics
 can be interpreted as an activated escape process induced by 
 noise fluctuations, and therefore the average residence time $<T_{res}>$
 in the attraction basin of the SM can be expressed in the following way
\begin{equation}
<T_{res}> \propto \exp{[W/R]}
\label{kramers}
\end{equation}
where $W$ represent a barrier that the orbit should overcome in order
to escape from the SM valley. As shown in Fig. \ref{fig:escape} 
the process can be indeed interpreted in terms of the Kramers
expression (\ref{kramers}) for $ 2\times 10^{-4} < R < 7 \times 10^{-4}$. 
It means that for $ R < W$ one should expect an intermittent behaviour,
while for $ R > W$ the dynamics of the orbit will be essentially diffusive, since 
the noise fluctuations are sufficient to drive the orbit always out
of the SM valley. These indications suggest that in order to observe 
a ``non transient'' or ``non intermittent'' LFF dynamics the amount of 
noise present in the system should be larger than $W$. 
Also in \cite{dfb} it has been clearly stated that the LK-equations 
with parameters tuned to reproduce the dynamics of a DFB laser
with $\alpha =3.4$ can give rise to stationary LFF only in presence of noise.

It is important to remark that for $\alpha=3.3$ the
experimentally measured variance of the noise $R=2.76 \times 10^{-3}$
is above the barrier $W=1.87 \times 10^{-3}$ found from the fit 
of the numerical $<T_{res}>$ with expression (\ref{kramers}), 
performed in the small noise range (see Fig. \ref{fig:escape}).
Moreover in the experiments we never observed relaxation of the dynamics towards a SM.

%%%%%%%%%%%%%%%%%%%%%%%%%%%%%%%%%%%%%%%%%%%%%%%%%%%%%%%%%%%%%%%%%%%%%%
\subsection{Lyapunov Analysis}
\label{sec:5.1}
%%%%%%%%%%%%%%%%%%%%%%%%%%%%%%%%%%%%%%%%%%%%%%%%%%%%%%%%%%%%%%%%%%%%%%

Also in the noisy case we have examined the degree of chaoticity
in the system by estimating the maximal Lyapunov exponent along
noisy orbits of the system. The role of noise is fundamental in
destabilizing the dynamics of the system and in rendering
the asymptotic dynamics chaotic.

\begin{figure}[h]
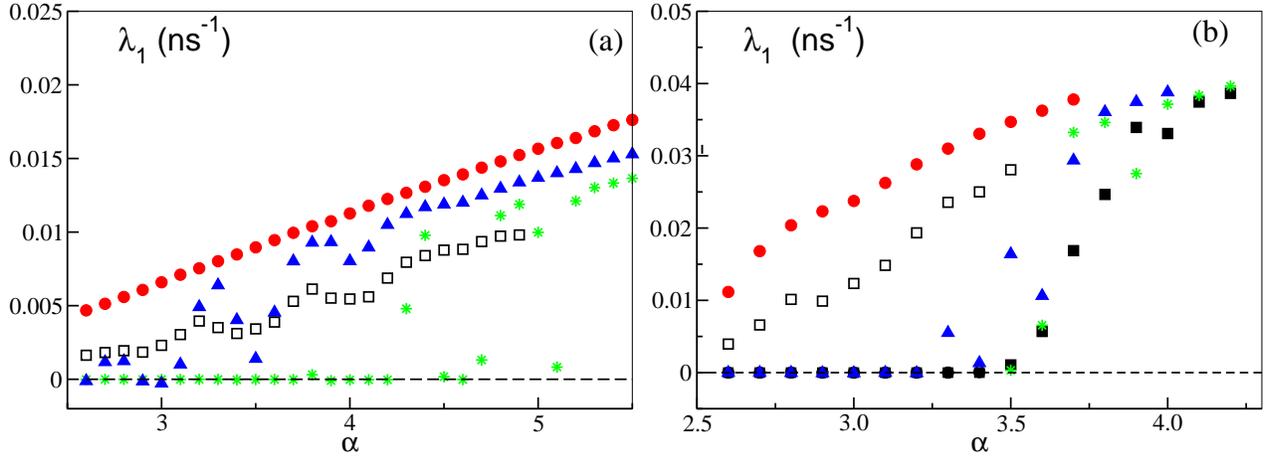

\centerline{
\includegraphics[draft=false, scale=.35,clip=true]{f10a.eps}
\includegraphics[draft=false, scale=.35,clip=true]{f10b.eps}
}
\caption{(color online) Maximal Lyapunov exponents $\lambda_1$ as a function
of $\alpha$ for $\mu=0.93$ (a) and $\mu=0.97$ (b) and for various noise amplitude
$R$: namely,  the data for $R=3 \times 10^{-5}$ are indicated by green asterisxs, 
those for $R=3 \times 10^{-4}$ by blue filled triangles, and the ones corresponding
to $R=3 \times 10^{-3}$ by red filled circles. The values estimated during
the transient dynamics in absence of noise are indicated by black empty squares,
while the asymptotic values for $R=0$ by black filled squares.
All the data refer to $k=0.25$, for the estimation of $\lambda_1$
in the noisy case one orbit has been followed for a time $t=3.63$~ms with time step
$\Delta t=0.363$~ps, while in the deterministic case the asymptotic results have also been
averaged over $M=10$ different initial conditions. For details on the estimation of the
transient Lyapunov exponents see the previous section \ref{sec:4.2}.
}
\label{fig:lyapnoise}
\end{figure}

 In particular, as shown in Fig. \ref{fig:lyapnoise}(a) we observe that at $\mu=0.93$ and $k=0.25$ the 
deterministic dynamics ($R=0$) is asymptotically stable in the
range $\alpha \le 4.4$, while the noisy dynamics becomes more
and more chaotic for increasing $R$. For the value 
$R=3.3 \times 10^{-3}$, close to the experimental one, the
dynamics is completely destabilized in the whole examined range $2.5 \le
\alpha \le 4.5$, while for smaller $R$-values the range of destabilization
is reduced. These results confirm the role of the noise in rendering the
LFF an asymptotic phenomenon. Moreover the maximal Lyapunov increeases
steadily with $\alpha$ at $R=3.3 \times 10^{-3}$. Similar findings
apply in the case $\mu =0.97$ (see Fig. \ref{fig:lyapnoise}(b)).

 The wild oscillations in the $<\lambda_1>$ values observable at level of noise
$ R < 10^{-4}$ reflect the stability properties of the SMs attracting
the asymptotic dynamics.

%%%%%%%%%%%%%%%%%%%%%%%%%%%%%%%%%%%%%%%%%%%%%%%%%%%%%%%%%%%%%%%%%%%%%%
\section{Comparison between experimental and numerical data}
\label{sec:6}
%%%%%%%%%%%%%%%%%%%%%%%%%%%%%%%%%%%%%%%%%%%%%%%%%%%%%%%%%%%%%%%%%%%%%%

 This Section will be devoted to a detailed comparison of numerical
versus experimental results with the aim to clarify if the deterministic
or noisy LK eqs are indeed able to reproduce the experimental findings.

%%%%%%%%%%%%%%%%%%%%%%%%%%%%%%%%%%%%%%%%%%%%%%%%%%%%%%%%%%%%%%%%%%%%%%
\subsection{Distributions of the field intensities}
\label{sec:6.1}
%%%%%%%%%%%%%%%%%%%%%%%%%%%%%%%%%%%%%%%%%%%%%%%%%%%%%%%%%%%%%%%%%%%%%%

\begin{figure}[h]
\includegraphics[draft=false, scale=.45,clip=true]{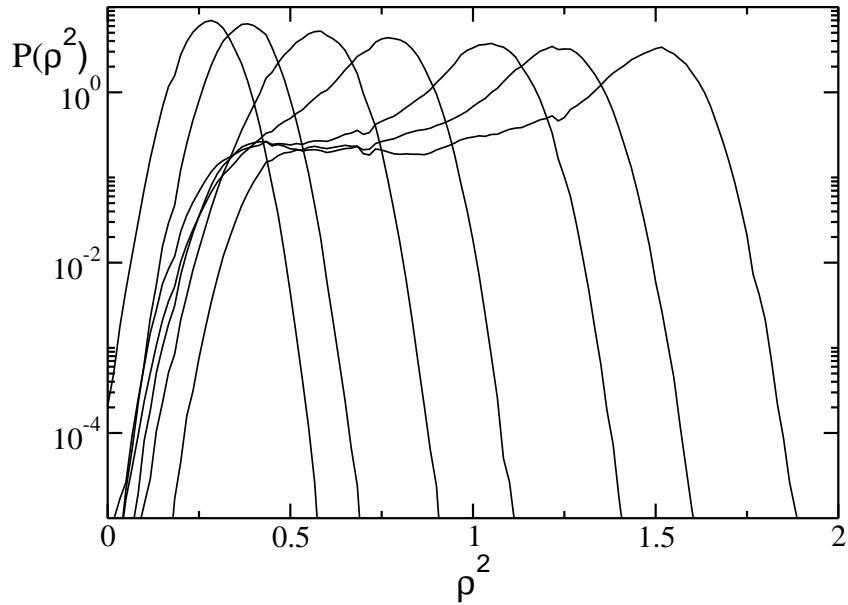}
\caption{Field intensities 
distributions $P(\rho^2)$ for the experimental signal filtered at 200 MHz.
The experimental intensities have been arbitrarly rescaled to match the 
corresponding average intensities obtained from the simulation of the noisy LK eqs 
at $\alpha=3.3$ and $k=0.35$ with noise variance $R=3.3\times 10^{-3}$. The data refer
from left to rigth to $I=2.48$ mA, 2.50, 2.54, 2.58, 2.64, 2.70, and 2.75.
Since $I_{th}=2.76$ these data correspond to $ 0.9 < \mu < 1.0$.
The first distribution has a large contribution from the Gaussian 
electronic noise, which also explains the negative $\rho^2$ values.
}
\label{fig:pdf_exp}
\end{figure}

 As a first indicator we have considered the distribution of the
field intensities $P(\rho^2)$, in particular in order to match the
experimental findings we consider the signal filtered at 200 MHz.

 The experimental results for the probability distribution functions (PDFs) 
of the field intensities $\rho^2$
are reported in Fig.~\ref{fig:pdf_exp} for various currents below
the solitary threshold value. It should be noticed that the amplitudes
$\rho^2$ have been rescaled in order to match the corresponding numerical values
for the noisy LK eqs with $\alpha=3.3$ and noise variance $R=3.3\times 10^{-3}$,
but that no arbitrarly shift have been applied to the data. 

 A peculiar characteristic of these data is that for increasing
pump current the PDFs become more and more asymmetric, revealing 
a peak at large intensities that shifts towards higher and higher 
$\rho^2$ values and a sort of plateau at smaller intensities.

\begin{figure}[h]
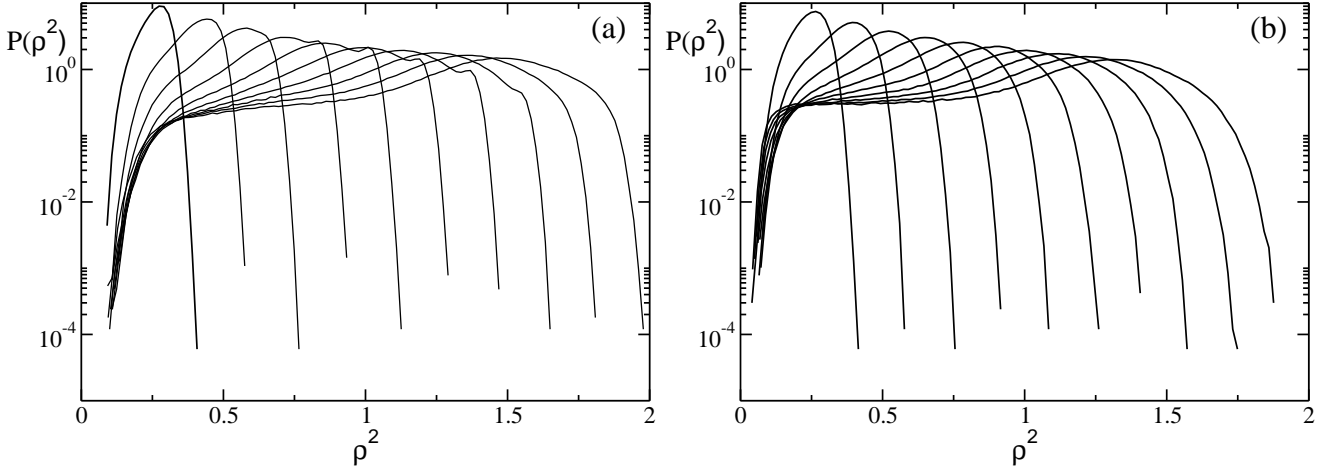

\centerline{
\includegraphics[draft=false, scale=.35,clip=true]{f12a.eps}
\includegraphics[draft=false, scale=.35,clip=true]{f12b.eps}
}
\caption{Field intensities 
distributions $P(\rho^2)$ for the numerical data obtained
by the integration of a noisy LK eqs. filtered at 200 MHz.
The data refer from left to rigth to $\mu=0.90$, 0.91, 0.92, 0.93, 0.94, 0.95,
0.96, 0.97, 0.98, and 0.99 for (a) $\alpha=3.3$ and (b) $\alpha=4.0$.
Both the sets of data correspond to $k=0.35$ and noise variance
$R=3\times 10^{-3}$.
}
\label{fig:pdf_sim.noise}
\end{figure}

\begin{figure}[h]
\includegraphics[draft=false, scale=.45,clip=true]{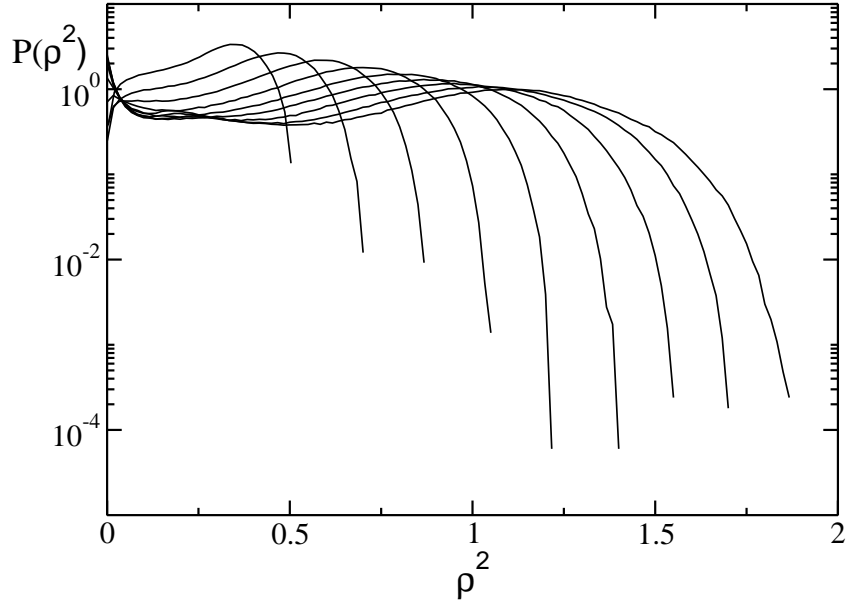}
\caption{Field intensities 
distributions $P(\rho^2)$ for the numerical data obtained
by the integration of a deterministic LK eqs. filtered at 200 MHz.
The data refer from left to rigth to $\mu=0.91$, 0.92, 0.93, 0.94, 0.95,
0.96, 0.97, 0.98, and 0.99 for $\alpha=5.0$  and $k=0.35$.
}
\label{fig:pdf_sim.det}
\end{figure}

 The corresponding PDFs are reported in Figs. \ref{fig:pdf_sim.noise}
and \ref{fig:pdf_sim.det} for data obtained from the integration of the 
noisy and deterministic LK eqs, respectively. A better agreement
between numerical and experimental findings is found for the noisy
dynamics with $\alpha \sim 3.3 - 4.0$ and $k=0.35$, with a noise variance
similar to the experimental one (namely, $R=3\times 10^{-3}$).
For the deterministic case (reported in Fig. \ref{fig:pdf_sim.det}
for $\alpha=5.0$ and $k=0.35$) a non zero tail at $\rho \sim 0$ is
observed even for $\mu \sim 1.0$, contrary to what observed for the
experimental data.

 These results indicate that it is necessary to include
the noise in the LK eqs to obtain a reasonable agreement with the
experiment, at least at the level of the intensities PDFs.
However, the numerical data seem unable to reproduce the narrow
peak present in the experimental ones at large intensities and for 
$\mu \to 1$. In the next sub-section
a further comparison will be performed to validate these preliminary
indications.

%%%%%%%%%%%%%%%%%%%%%%%%%%%%%%%%%%%%%%%%%%%%%%%%%%%%%%%%%%%%%%%%%%%%%%
\subsection{Average values of the LFF times}
\label{sec:6.2}
%%%%%%%%%%%%%%%%%%%%%%%%%%%%%%%%%%%%%%%%%%%%%%%%%%%%%%%%%%%%%%%%%%%%%%

 We will first compare the experimental and numerical measurements of the
average times between two consecutive drops of the field intensities $<T_{LFF}>$. 
They have been evaluated
in two (consistent) ways: both from a direct measurement of periods between
threshold crossing, and from the Fourier power spectrum of the
temporal signal $\rho^2(t)$.

 The direct measurements of the $T_{LFF}$ from the time trace of $\rho^2(t)$
have been performed by defining two thresholds $\Gamma_1 < \Gamma_2$ and
by identifing two consecutive time crossing of $\Gamma_1$, provided that 
in the intermediate time the signal has overcome the threshold 
$\Gamma_2$ at least once. The thresholds have been defined as
$\Gamma_1= <\rho^2> - 2*STD$ and $\Gamma_2= <\rho^2> + STD/2$,
where $< \cdot> $ and STD indicate the average and the 
standard deviation of the signal itself.

 The measurement of $<T_{LFF}>$ in terms of the power spectrum has
been obtained by considering the power spectrum $S(\omega)$ of $\rho^2(t)$ 
and by evaluating the position $\omega_{M}$ of the peak with the highest
frequency, then $<T_{LFF}> = 2 \pi / \omega_{M}$. As already mentioned the
two estimations are generally in very good agreement.

\begin{figure}[h]
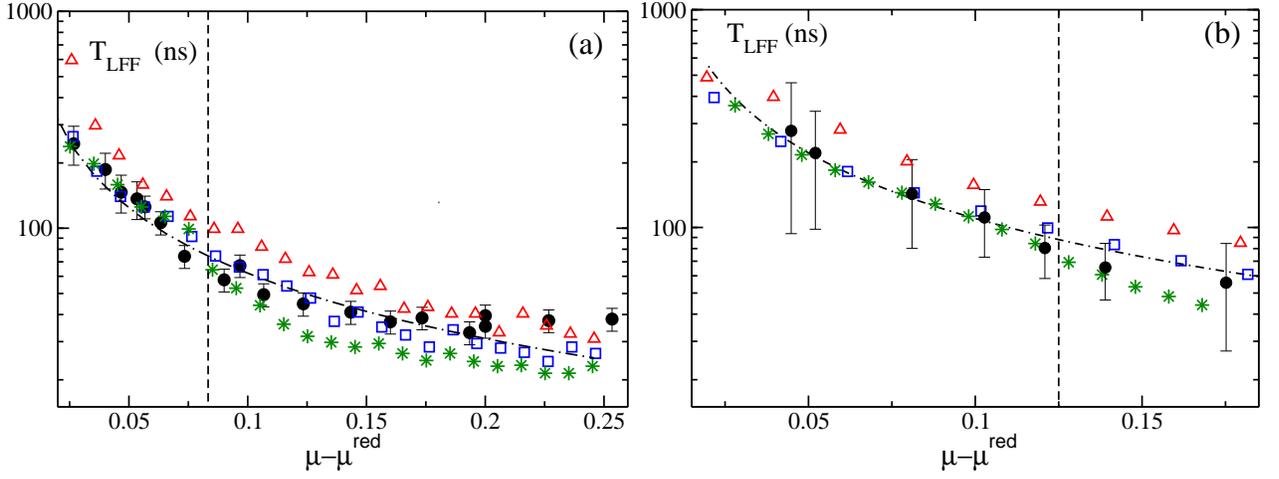

\centerline{
\includegraphics[draft=false, scale=.35,clip=true]{f14a.eps}
\includegraphics[draft=false, scale=.35,clip=true]{f14b.eps}
}
\caption{(Color online) 
Average LFF times $<T_{LFF}>$ as a function of the pump parameter
$\mu - \mu^{red}$: black filled circles refer to experimental data, while the other symbols to the
results obtained from the integration of the LK eqs. 
In the two figures are reported two different sets
of experimental measures and the associated numerical data refer
to $k=0.25$ (a) and $k=0.35$ (b).
In particular, red empty triangles correspond to the evolution of the noisy LK at 
$\alpha=3.3$ (with (a) $\mu^{red}=0.914$ and (b) $\mu^{red}=$0.880)
and blue empty squares to $\alpha=4.0$ (with $\mu^{red}=0.913$ and 0.878, resp. for (a) and (b)).
In both cases $R=3\times 10^{-3}$. The green stars denote the data of the deterministic LK eqs 
for $\alpha=5.0$ (in this case $\mu^{red}=0.915$ and 0.882, resp. for (a) and (b)). 
For the experimental measures $\mu^{red}=0.916$ in (a) and 0.875 in (b).
The vertical dashed lines indicates the position of the solitary threshold for the experimental
data, while the dash-dotted lines represent the decay $c/(\mu -\mu^{red})$, with
$c=6.2$ ns and 11 ns in (a) and (b), respectively. $\mu^{red}$ is defined as the ratio
$I^{red}_{th}/I_{th}$.
}
\label{fig:T_LFF}
\end{figure}

In Fig. \ref{fig:T_LFF} the average times $<T_{LFF}>$ are reported for two 
different sets of experimental measurements as a function of the pump parameter
$\mu$ and compared with numerical data. In Fig. \ref{fig:T_LFF} (a) are
reported the experimental findings already shown in \cite{romanelli},
the estimation of $T_{LFF}$ have been performed both by direct inspection
of the signal and via the first zero of the 
autocorrelation function (this second method corresponds
to an evaluation from the Fourier power spectrum). 
The numerical data
have been obtained with the 2 methods outlined above for $k=0.25$
for both noisy and deterministic LK eqs. 
In Fig. \ref{fig:T_LFF} (b) a new set of experimental
data is reported and compared with simulation results for $k=0.35$,
in this case all the data have been obtained by the method of the thresholds.
From the figures it is clear
that a reasonably good agreement between experimental and numerical
data is observed for the deterministic case only for $\alpha=5$ 
(results for smaller $\alpha$-values, obtained during the transient
preceeding the stable phase, are not shown but they exhibits a worst 
agreement with experimental findings) and for the noisy dynamics
for $\alpha=4.0$. 

\begin{figure}[h]
\centerline{
\includegraphics[draft=false, scale=.35,clip=true]{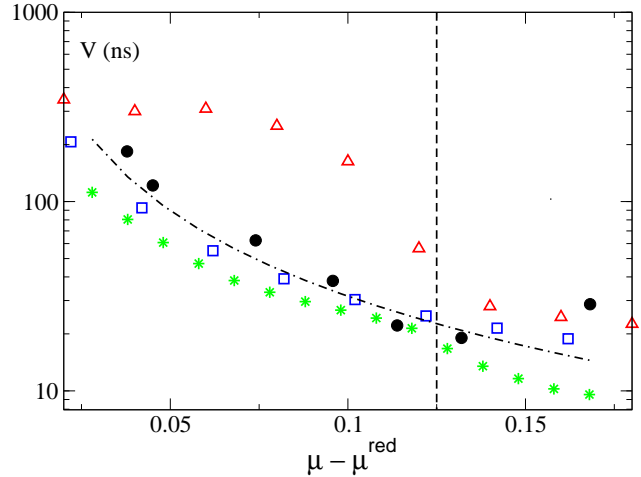}
}
\caption{(Color online) 
Standard deviation of the LFF times $V$ as a function of the pump parameter
$\mu -\mu^{red}$: the symbols are the same as those reported in 
Fig.~\ref{fig:T_LFF} (b). The dash-dotted line indicates the power-law decay $1/(\mu -\mu^{red})^{3/2}$. 
}
\label{fig:T_V}
\end{figure}

A more detailed analysis can be obtained by considering not only the average 
values of the LFF times, but also the associated standard deviation $V$.
This quantity reported in Fig.~\ref{fig:T_V} exibits a clear decrease
with $\mu$ by approaching the solitary treshold, indicating a
modification of the observed dynamics that tends to be more ``regular''.
Also in this case the comparison of experimental and numerical data
suggests that the best agreement is again attained with the noisy dynamics at
$\alpha=4.0$.

At this stage of the comparison we can sketch some preliminary
conclusions: the LK eqs are able to reproduce reasonably well
the experimental data for the VCSEL below the solitary threshold
both in the deterministic case  and in the noisy situation.
However in the deterministic case a quite large value of the linewidth enhancement
factor (with respect to the experimentally measured one) is required.
A more detailed comparison will be possible by considering the PDFs
of the $T_{LFF}$.

%%%%%%%%%%%%%%%%%%%%%%%%%%%%%%%%%%%%%%%%%%%%%%%%%%%%%%%%%%%%%%%%%%%%%%
\subsection{Distributions of the LFF times}
\label{sec:6.3}
%%%%%%%%%%%%%%%%%%%%%%%%%%%%%%%%%%%%%%%%%%%%%%%%%%%%%%%%%%%%%%%%%%%%%%

\begin{figure}[h]
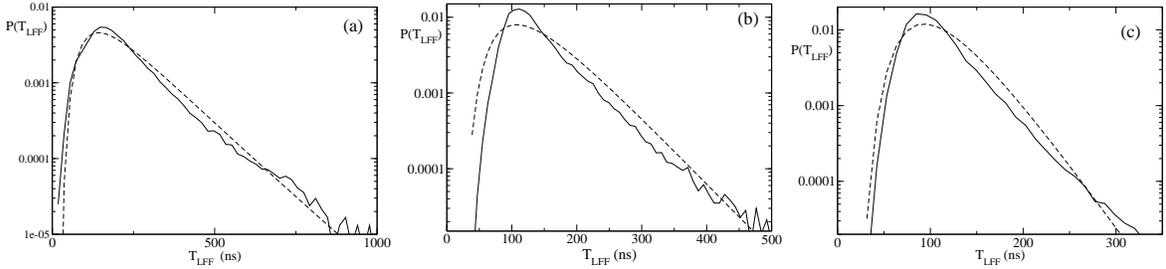

\centerline{
\includegraphics[draft=false, scale=.2,clip=true]{f15a}
\includegraphics[draft=false, scale=.2,clip=true]{f15b}
\includegraphics[draft=false, scale=.2,clip=true]{f15c}
}
\caption{Probability density distributions of the $T_{LFF}$.
Solid lines refer to experimental data, the dashed ones
to the Inverse Gaussian Distribution (\ref{inv_gau}) with the average 
and standard deviation corresponding to the experimental ones.
(a) $I=2.56$ mA, (b) $I=2.64$ mA, and (c) $I=2.70$ mA
}
\label{fig:pdf_LFF}
\end{figure}

In this sub-section we will examine the whole distribution of the
$T_{LFF}$ in more details. Considering the experimental
data, we observe that all the measured PDFs obtained for different
pump currents reveal an exponential-like tail at long times and
a rapid drop at short times (as shown in Fig. \ref{fig:pdf_LFF}).
These results are in agreement with those reported in \cite{sukov}
for a single-transverse-mode semiconductor laser in proximity of $I_{th}$.

The typical dynamics corresponding to a LFF can be summarized as follows: 
a sudden drop of the intensity is followed
by a steady increase of $\rho^2$, associated to fluctuations
of the intensity, until a certain threshold is reached and the
intensity is reset to its initial value and restart
with the same ``sysiphus cycle'' \cite{sano}. This behaviour 
and the observed shapes of the PDFs suggest that the dynamics of
the intensities can be modelized in terms of a Brownian motion plus drift.
In other words, by denoting with $x(t)$ the intensity, 
an effective equation of the following type can be written
to reproduce its dynamical behaviour:
\begin{equation}
\dot x(t) = \eta + \sigma \xi(t)
\label{bm_drift}
\end{equation}
with initial condition $x(0)=x_0$, where $\xi(t)$ is a Gaussian noise term with zero average
and unitary variance, $\eta$ represents the drift, $\sigma$ is the noise strength.
Within this framework the average first passage time to reach a fixed threshold $\Gamma$
is simply given by $\tau= (\Gamma - x_0)/\eta$ , while the corresponding standard
deviation is $V = [\sqrt{(\Gamma - x_0)} \sigma] / \eta^{3/2}$~\cite{tuckwell}.
A resonable assumption would be that $\eta$ is directly proportional to the pump parameter $(\mu - \mu^{red})$,
(where $\mu^{red}=I^{red}_{th}/I_{th}$ is the rescaled pump current value at the reduced threshold)
and by further  assuming that the threshold $\Gamma$ is independent on the pump current this would imply that 
\begin{equation}
 <T_{LFF}> = \frac{c}{(\mu - \mu^{red})} \qquad, \quad {\rm and} \quad
 V_{LFF} = \frac{c}{(\mu - \mu^{red})^{3/2}}
\label{teo_estim}
\end{equation}
these dependences are indeed quite well verified for the experimental data above the solitary
threshold as shown in Figs~\ref{fig:T_LFF} and \ref{fig:T_V}.

For the simple model introduced by eq. (\ref{bm_drift}),
the PDF of the first passage times is the so-called Inverse Gaussian Distribution~\cite{gaussiana}:
\begin{equation}
P(T)=\frac{\tau}{\sqrt{2 \pi \gamma T^3}}
{\rm e}^{-(T-\tau)^2/(2 \gamma T)}
\label{inv_gau}
\end{equation}
where $\gamma= V^2/\tau$. A comparison of this expression with the experimentally
measured $P(T_{LFF})$ is reported in Fig. \ref{fig:pdf_LFF}. The good agreement
suggests that the ``sysiphus cycles'' can be due to few elementary ingredients:
a stochastic motion subjected to a drift plus a reset mechanism once the
intensity has overcome a certain threshold.

\begin{figure}[h]
\includegraphics[draft=false, scale=.4,angle=-90,clip=true]{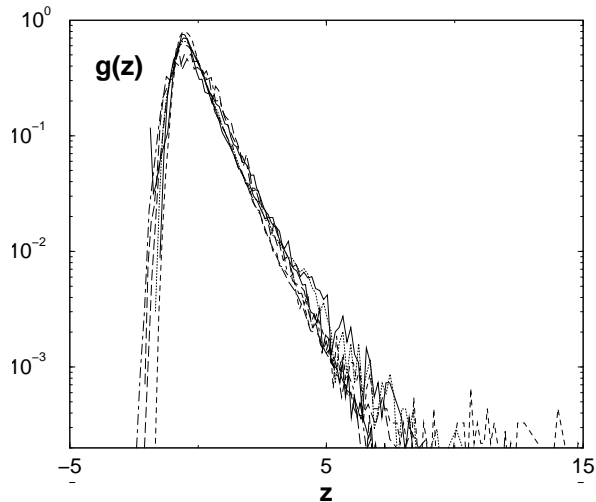}
\caption{Rescaled probability density distributions $g(z)=V*P(T_{LFF})$
as a function of $z=(T_{LFF}-<T_{LFF}>)/V$. The curves refer to
$I=2.54$ mA, 2.56 mA, 2.64 mA, 2.70 mA, 2.75 mA, 2.80 mA,
and 2.90 mA.}
\label{fig:rescaled_exp}
\end{figure}

 A way of rewriting the distribution (\ref{inv_gau}) in a more compact form
as a function of only one parameter, the so-called 
coefficient of variation $\delta=V/\tau$ (i.e., the ratio between 
the standard deviation $V$ and the mean $\tau$), is to rescale
the time as $z = (T- \tau)/V$ and the PDFs as $g(z)=V*P(T)$. This
procedure leads to the following expression:
\begin{equation}
g(z)=\frac{1}{\sqrt{2 \pi (\delta z+1)^3}}
{\rm e}^{-z^2/2(\delta z+1)} \quad .
\label{inv_gau_res}
\end{equation}
It is clear that all the PDFs will coincide, once rescaled in this way, if
the coefficient of variation
$\delta$ has the same value for all the considered pump currents.
However this is not the case and indeed we measured values of $\delta$
in the range $[0.28:0.66]$ for $ I < I_{th}$, nonetheless if
we report in a single graph all these curves the overall matching is
very good, as shown in Fig. \ref{fig:rescaled_exp}.

\begin{figure}[h]
\centerline{
\includegraphics[draft=false,scale=.45,angle=-90,clip=true]{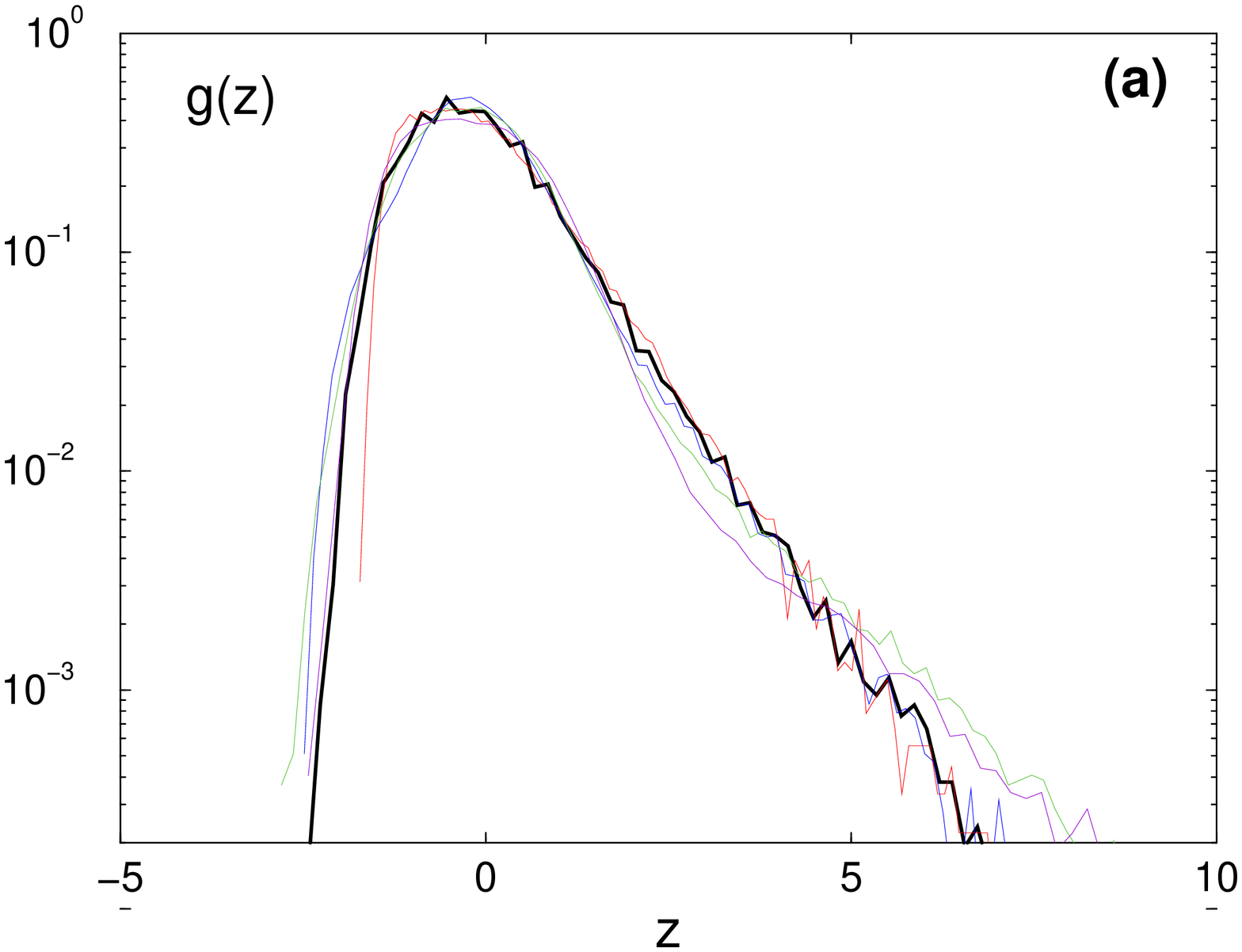}
\includegraphics[draft=false,scale=.45,angle=-90,clip=true]{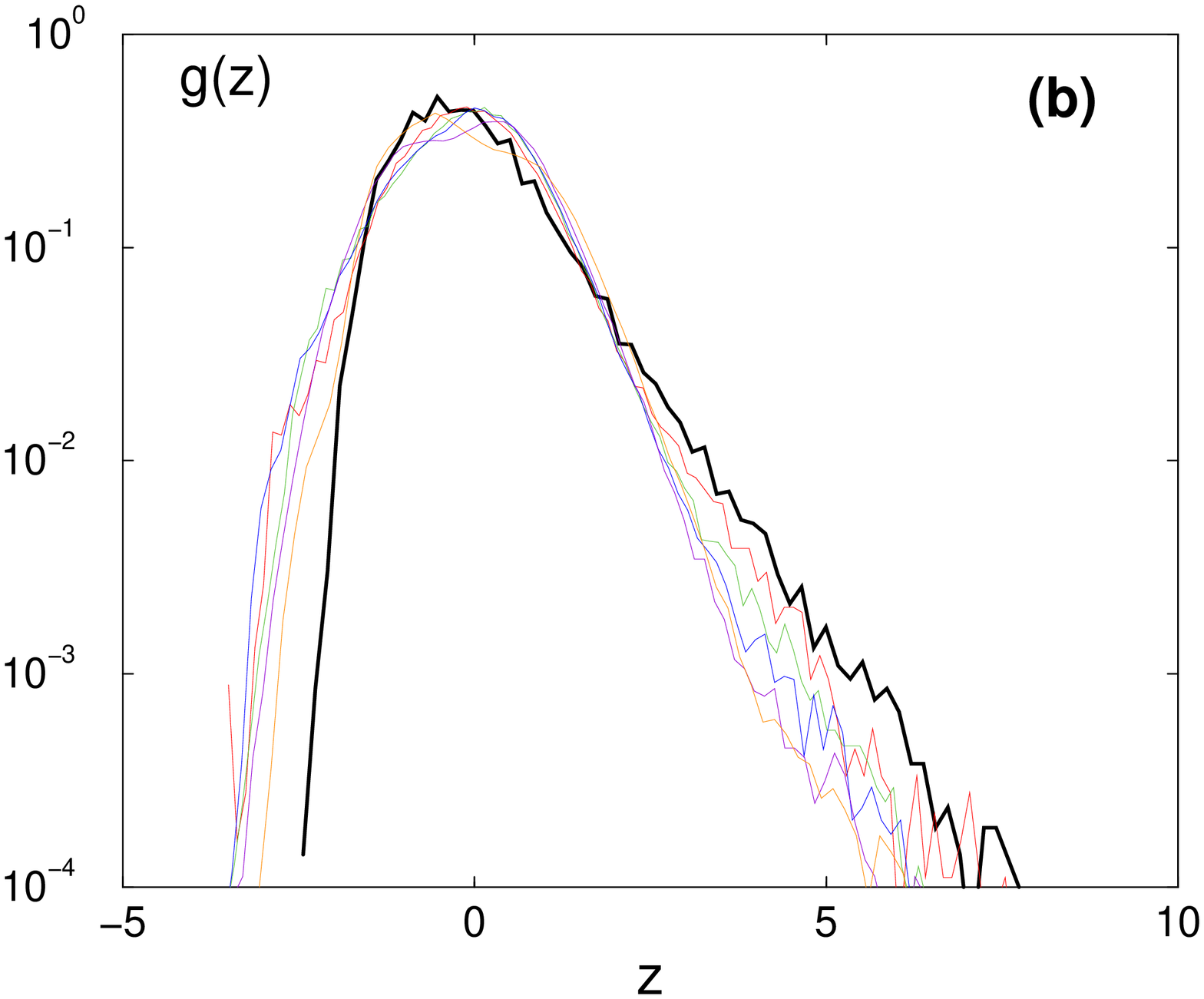}
}
\caption{(Color line) Rescaled probability density distributions $g(z)=V*P(T_{LFF})$
as a function of $z=(T_{LFF}-<T_{LFF}>)/V$. The thick solid curve refer to
experimental data for $I=2.75$ mA (corresponding to $\mu \simeq 1$),
the other ones to numerical findings: (a) data obtained for
the noisy LK eqs for $\alpha=4.0$, $k=0.35$ with variance $R=3 \times
10^{-3}$ and corresponding to $\mu=0.90$ (red), 0.94 (blue), 
0.98 (green), and 1.02 (violet); (b)
data for the the deterministic LK eqs. for $\alpha=5.0$, $k=0.35$
corresponding to $\mu=0.92$ (red) , 0.94 (blue), 0.96 (green), 0.98 (violet),
and 1.00 (orange). 
}
\label{fig:rescaled_sim}
\end{figure}

Let us finally compare these distributions $g(z)$ with the corresponding ones
obtained from direct simulations of the LK eqs.. As one can see from 
Fig. \ref{fig:rescaled_sim} the agreement is good for the data obtained
from the simulation of the noisy LK eqs at $\alpha=4.0$, while it
is worse for the deterministic LK eqs at $\alpha=5.0$.

%%%%%%%%%%%%%%%%%%%%%%%%%%%%%%%%%%%%%%%%%%%%%%%%%%%%%%%%%%%%%%%%%%%%%%%%%%%
\section{Conclusions}
\label{sec:7}
%%%%%%%%%%%%%%%%%%%%%%%%%%%%%%%%%%%%%%%%%%%%%%%%%%%%%%%%%%%%%%%%%%%%%%%%%%%
We presented a detailed experimental and numerical study of a semiconductor laser with optical feedback.
The choice of a Vertical Cavity Laser pumped close to its threshold, together with a polarized optical
feedback, assures a great control over the possibility of lasing action of other orders longitudinal
and/or transverse modes than the fundamental one, and of the activation of the other polarization. In such a
way, the description of the system using the Lang-Kobayashi model is well justified and it allows for a meaningful
comparison with the experimental data. The analysis has been performed with particular regard to the LFF regime,
where the model has been numerically integrated using parameters carefully measured in the laser sample used for the measurements.

The comparison of the the measurements carried out in the VCSEL with polarized optical feedback
with the predictions of the deterministic LK model suggest that in the examined range of parameters the 
dynamics of the model is characterized by a chaotic transient leading to stable ECMs with high gain. 
The transient duration increases (and possibly diverges) 
with increasing values of the rescaled pump current $\mu$ and of the linewidth enhancement factor $\alpha$.
We have not found evidence of periodic or quasi-periodic asymptotic attractors, as instead 
reported in \cite{david}, this can be due to the $\alpha$-range examined 
in the present paper (namely $2.4 \le \alpha \le 5.5$), since these solutions become relevant 
for the dynamics only for $\alpha > 5$ (as stated in \cite{david}).

However, a stationary LFF dynamics with characteristics similar to those measured experimentally,
can be obtained for realistic values of the $\alpha$ parameter (namely, $\alpha \sim 3 - 4$)
only via the introduction of an additive noise term in the LK equations. The role of noise in determining the
statistics and the nature of the dropout events has been previously examined in \cite{hk,hohl}, but
in the present paper we have clarified that the LFF dynamics can be interpreted at a 
first level of approximation as a biased Brownian motion towards a threshold with a reset mechanism.

%%%%%%%%%%%%%%%%%%%%%%%%%%%%%%%%%%%%%%%%%%%%%%%%%%%%%%%%%%%%%%%%% 
\acknowledgments
%%%%%%%%%%%%%%%%%%%%%%%%%%%%%%%%%%%%%%%%%%%%%%%%%%%%%%%%%%%%%%%%%
We acknowledge useful discussions with M. B\"ar, S. Yanchuk,
S. Lepri, and M. Wolfrum. Two of us (G.G. and F.M.) 
thanks C. Piovesan for its effective support.

%%%%%%%%%%%%%%%%%%%%%%%%%%%%%%%%%%%%%%%%%%%%%%%%%%%%%%%%%%%%%%%%%%%%%%%

%%%%%%%%%%%%%%%%%%%%%%%%%%%%%%%%%%%%%%%%%%%%%%%%%%%%%%%%%%%%%%%%%%%%%%%

\end{document}